\UseRawInputEncoding
\documentclass[aps,onecolumn,superscriptaddress,groupedaddress,nofootinbib]{revtex4}

\pdfoutput=1
\usepackage{dcolumn}
\usepackage{bm}
\usepackage{mathrsfs}
\usepackage{hyperref}
\usepackage{overpic}

\usepackage{amsmath}
\usepackage{amssymb}
\usepackage{bm}
\usepackage{color}
\usepackage{float}
\usepackage{dcolumn}
\usepackage{multirow}
\usepackage{changepage}
\usepackage{enumerate}
\usepackage{float}
\usepackage{subfloat}
\usepackage{subfig}
\usepackage{graphicx}
\usepackage{ulem}
\hypersetup{pdftex,colorlinks=true,linkcolor=blue,citecolor=red,menucolor=black,urlcolor=blue,filecolor=blue}

\begin{document}

\title{Einstein-de Haas effect: a bridge linking mechanics, magnetism, and topology}

\author{Xin Nie}
\author{Dao-Xin Yao}
\email{yaodaox@mail.sysu.edu.cn}
\affiliation{Guangdong Provincial Key Laboratory of Magnetoelectric Physics and Devices, State Key Laboratory of Optoelectronic Materials and Technologies, Center for Neutron Science and Technology, School of Physics, Sun Yat-Sen University, Guangzhou, 510275, China}
\begin{abstract} 
The Einstein-de Haas (EdH) effect is a fascinating phenomenon that links mechanics and magnetism. Despite being discovered over a century ago, it remains significant in contemporary science, particularly within the fields of spintronics and ultrafast magnetism. Recent predictions suggest that the EdH effect may be realized in topological magnon systems, potentially leading to even richer properties. In this perspective, we introduce recent advancements in the EdH effect and discuss its developments in three key aspects: the microscopic mechanism, its manifestation in topological systems, and chirality-selective magnon-phonon coupling. Our discussions aim to inspire further explorations of the EdH effect and highlight its promising applications in different areas.
\end{abstract}

\maketitle

\begin{figure*}[!h]
\centering
\subfloat[]{ \includegraphics[width=3 in]{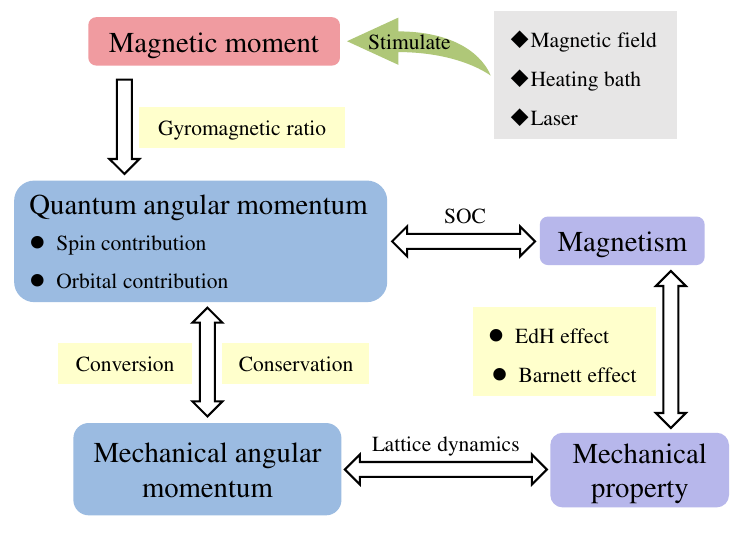}\label{Fundamentals}}
\subfloat[]{ \includegraphics[width=1.15 in]{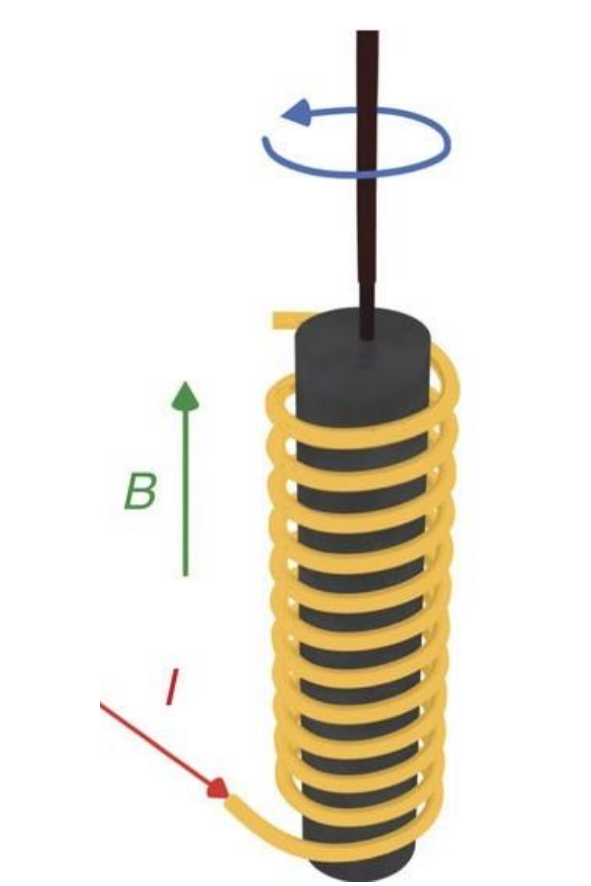}\label{EdH_p}}
\subfloat[]{ \includegraphics[width=3.1 in]{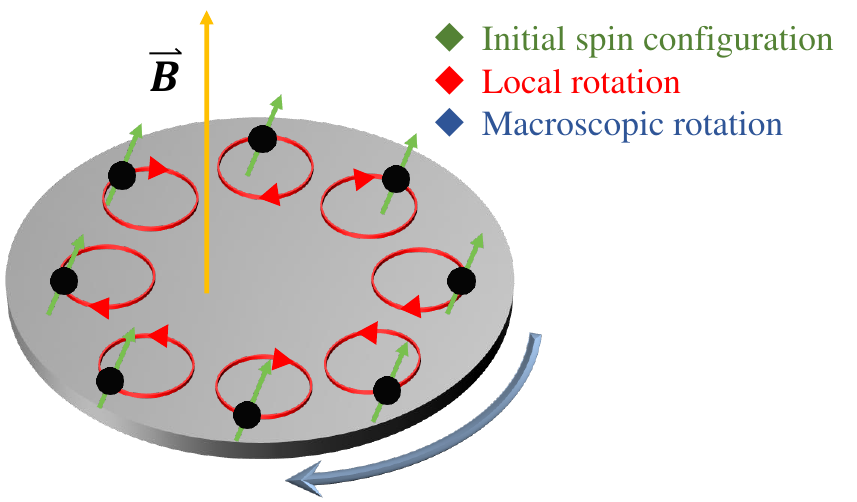}\label{SRC}}
\caption{(a) Fundamentals of the EdH effect. (b) EdH experiment \cite{Ganzhorn2016}. (c) Spin-rotation coupling mechanism \cite{Nie2024}. 
}
\end{figure*}

The EdH effect, which establishes that changes in magnetization induce mechanical rotation, was experimentally demonstrated in 1915 \cite{Einstein1915}, independently of Richardson's earlier conceptual prediction related to this aspect \cite{PhysRevSeriesI.26.248}. In the original experiment, a cylindrical ferromagnet was wrapped with a wire, as illustrated in Fig. \ref{EdH_p}. When an electric current is injected into the wire, creating a magnetic field, the magnet is observed to rotate. This rotation occurs because the magnetic field alters the orientations of electron magnetic moments. The angular momentum of electrons relates to their magnetic moments through the gyromagnetic ratio (defined as the ratio of magnetic moment to angular momentum), and also changes as a result. Thus, the magnet rotates to conserve the total angular momentum in the z-direction ($\vec{B}\parallel z$). This experiment is called the EdH experiment. Soon after that, Barnett discovered the inverse of the EdH effect, showing that the rotational motion of a magnet influenced its magnetization \cite{PhysRev.6.239}. Both gyromagnetic effects are rooted in the conservation of angular momentum between electrons and mechanical rotation. However, the underlying mechanism about how the angular momentum of electrons is converted into that of macroscopic rotation has been elusive, primarily due to the intricate interplay among three fundamental degrees of freedom (see Fig. \ref{Fundamentals}): electron spin, orbital motion, and lattice dynamics. Numerous theoretical studies have focused on the coupled dynamics of some of these degrees of freedom, rather than all of them. In the case of an individual spin embedded in lattice, researchers typically connect the two via the spin-phonon interaction \cite{PhysRevB.92.024421, PhysRevB.72.094426, PhysRevB.97.174403}, which is derived from magnetic anisotropy. Moreover, electron-phonon interactions have been utilized to explain the EdH effect. A novel quasiparticle, termed ``angulon'', has recently been introduced to account for the interactions among electron spin, orbital motion, and phonon degrees of freedom \cite{PhysRevB.99.064428}. While these quantum models provide valuable insights into the angular momentum transfer from spins or electrons to atoms, they do not incorporate the angular momentum associated with rigid-body rotation. In classical frameworks, this transfer is commonly attributed to either magnetocrystalline anisotropy or magnetoelastic coupling. Researchers have applied magneto-molecular dynamics to simulate the angular momentum transfer between spins and lattice \cite{PhysRevB.93.060402, DEDNAM2022111359, AMANN2019217, PhysRevB.78.024434}. However, these theories fail to explain specific processes for the angular momentum transfer. Furthermore, both quantum and classical theories lack experimental validation.

Recent advances in ultrafast demagnetization experiments have significantly enhanced our understanding of the EdH effect's microscopic origins \cite{Dornes2019, Koopmans2010, Stamm2007, PhysRevLett.115.217204}: upon femtosecond laser excitation, the incident photon energy is immediately absorbed by electrons, propelling them to higher energy states within their atomic orbitals. In the presence of spin-orbit coupling, the spins begin to flip, and their angular momentum changes. It is noteworthy that the angular momentum brought by the photon during this process is extremely small and is negligible. Since the electronic orbit is incapable of accommodating the majority of the angular momentum lost by the spin, the spin redirects its angular momentum to ions in the lattice. In this way, a physical picture of ultrafast demagnetization is established, in which the primary demagnetization occurs through the transfer of angular momentum from spins to lattice. This channel elucidates why the EdH effect can be observed in most transition metals where the orbital angular momentum is readily frozen by the crystal field. Therefore, the next crucial step in elucidating the EdH effect is to clarify the dynamics of angular momentum transfer from spins to phonons or the lattice. 

In previous research, the EdH effect, especially the Barnett effect, was often explained by a phenomenological model that incorporates the coupling between spin and rigid-body rotation, represented as $\bm{S}\cdot\bm{\Omega}$, where $\bm{S}$ denotes the spin angular momentum and $\bm{\Omega}$ denotes the angular velocity of the rigid-body rotation. Although the model satisfies the conservation of angular momentum, it poses a subtle question at the microscopic level: Whether the spin angular momentum can be directly transferred to the macroscopic rotation without traversing phonons? To answer the question, we envision such a scenario \cite{PhysRevB.101.104402} in which a magnon is removed from the magnetic system and the reduced angular momentum is fully converted into macroscopic rotation. Based on the conservation of angular momentum, the rotational angular velocity of the magnet can be expressed as $\bm{\Omega} = \hbar\bm{S}/I$, with the Planck's constant $\hbar$, the spin $\bm{S}$, and the moment of inertia $I$. Therefore, we can deduce the kinetic energy of lattice $\frac{1}{2}I{\Omega}^2$, which is much lower than the energy needed to flip a spin, $\hbar w$, where $w$ is the magnon frequency. This discrepancy indicates that there should be an intermediary involved in the process of exchanging angular momentum and energy, which is most likely phonons. This view is further supported by the recent ultrafast experiment \cite{Tauchert2022}.

Researchers have characterized the phonon-mediated mechanism by exploiting spin--rotation coupling \cite{Nie2024}, which comprises local and macroscopic rotations, as shown in Fig. \ref{SRC}. When a stationary elastic magnet is subjected to a magnetic field misaligned with its initial ferromagnetic configuration, its spin angular momentum undergoes a change and is subsequently transferred to the atoms. This transfer induces local rotations of the atoms around their equilibrium positions, thereby forming circularly polarized phonons. The angular momentum is then transferred to the entire lattice, resulting in the macroscopic EdH rotation. Based on this picture, a theoretical model for the spin-lattice coupling is constructed--changes in magnetization generate a force on the atom, while the evolution of atomic displacement produces a torque acting on the spin. By numerically solving the coupled dynamics, the transfers of angular momentum and energy are quantitatively analyzed, and the timescale for angular momentum transfer from spins to lattice is found to match observations from demagnetization experiments \cite{10.1063/1.4958846}. Undoubtedly, more direct and compelling evidence for this mechanism would be the calculation of phonon angular momentum. However, Ref. \cite{Nie2024} did not perform a direct calculation of phonon angular momentum; instead, it provided indirect support for its existence. This limitation arises from the inherent challenges associated with continuum field theory, where distinguishing local rotations from macroscopic dynamics remains a complex task. Additionally, the transfer of angular momentum from spins to phonons occurs on a timescale that is considerably shorter than that from phonons to the lattice, which further complicates theoretical calculation and experimental detection of the phonon angular momentum. Moreover, an important question emerges for further exploration: how is the angular momentum of lattice vibrations transferred to the macroscopic rotation of the sample? This question remains unresolved.

\textit{Experimental progress.} The EdH technique has been widely employed for measuring the g-factor \cite{RevModPhys.34.102} of materials since its discovery because of its superior accuracy compared to electron-spin and ferromagnetic resonance methods. Over the past two decades, the applications of EdH experiments have broadened significantly. Firstly, the experimental systems have evolved from the macroscopic scale to the microscopic scale; secondly, the range of experimental materials has expanded from conventional ferromagnets to include antiferromagnets \cite{Zong2023}. Moreover, both the quantum EdH experiment \cite{Ganzhorn2016} and the ultrafast Barnett experiment \cite{Davies2024} have been successfully realized. 

\textit{EdH effect in biology.} As a fundamental concept, the EdH effect has garnered attention not only in physics but also in biology, where it manifests as atomic spin-mechanical coupling. In nature, many organisms can sense the Earth's magnetic field for navigation or migration, a phenomenon referred to as magnetoreception. Despite its importance, the biophysical mechanisms of magnetoreception remain unknown. In 2016, researchers reported a magnetic receptor (Drosophila CG8198, MagR) with a rod-like protein complex that assisted animals in guiding their magnetic orientation \cite{Qin2016}. Subsequently, physicists utilized atomic spin-mechanical interaction to demonstrate the effective alignment of the protein's magnetic moment with the Earth's magnetic field at room temperature \cite{PhysRevE.97.042409}. This finding clarified an ongoing debate concerning the thermal behaviors of MagR \cite{10.7554/eLife.17210}. Indeed, atomic-scale EdH effects have already been explored theoretically, with studies predicting that both a two-dysprosium-atom system trapped in a spherically symmetric harmonic potential \cite{Gorecki2016MakingTD} and an $O_{2}$ dimer system modeled using a noncollinear tight-binding approach \cite{10.1063/1.5092223} can exhibit this effect.

\begin{figure*}[!h]
\centering
\subfloat[]{ \includegraphics[width=3.8 in]{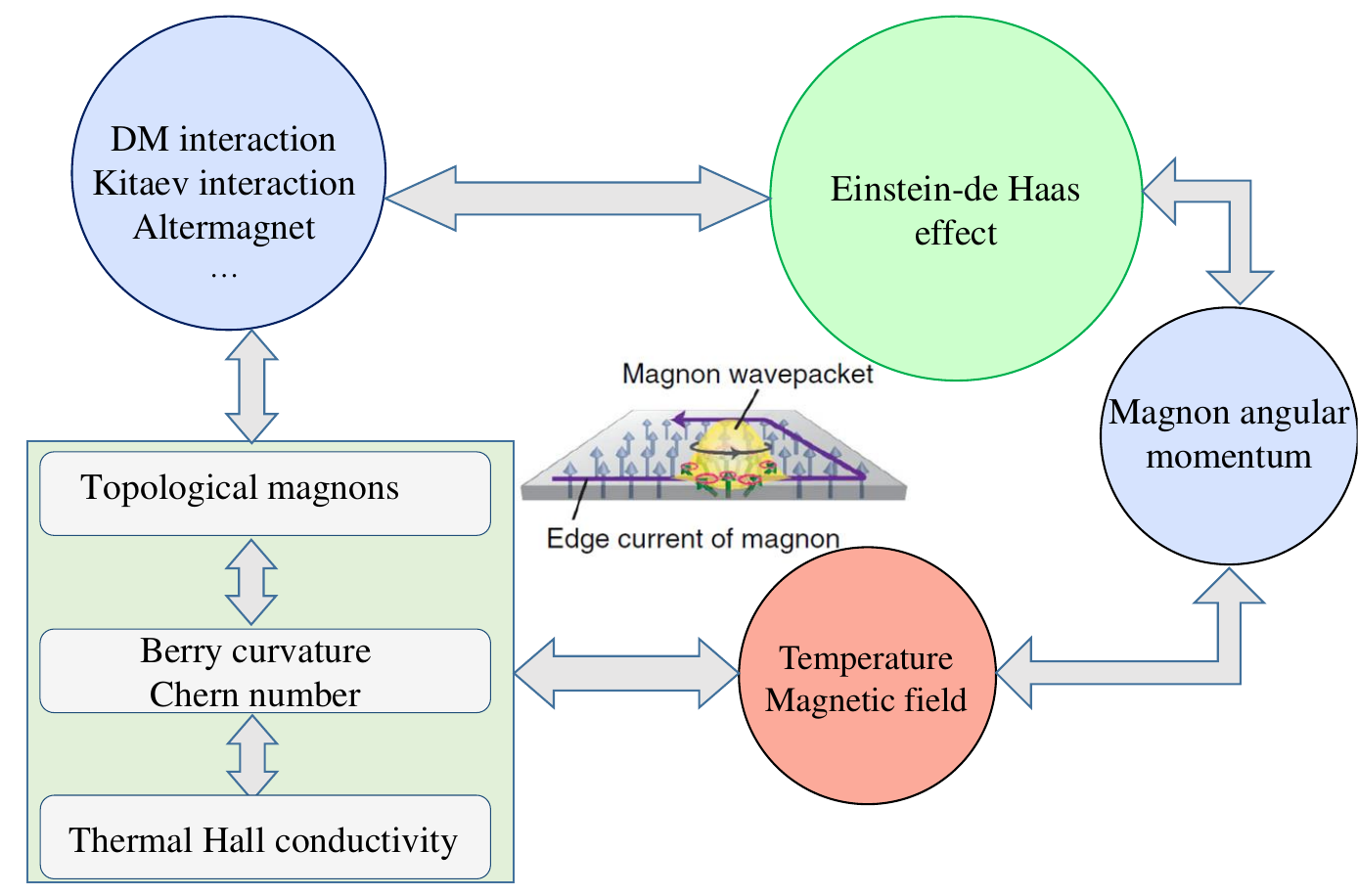}\label{two rotations}}
\quad
\subfloat[]{ \includegraphics[width=3 in]{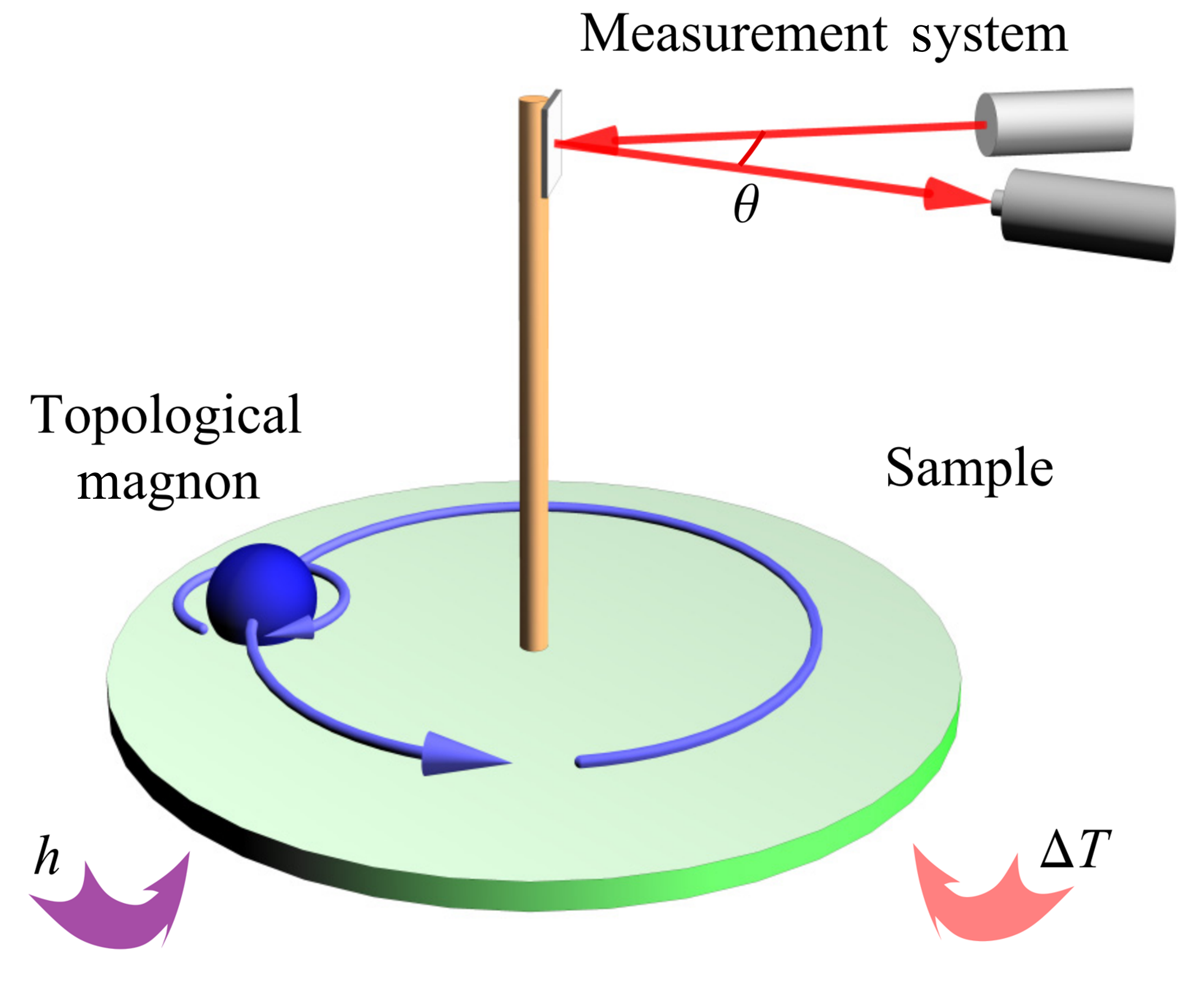}\label{exp}}
\caption{EdH effect of toplogical magnons \cite{PhysRevResearch.3.023248, PhysRevLett.106.197202}. (a) Theoretical mechanism. (b) Experimental design. 
}
\end{figure*}

\textit{EdH effect of topological magnons.} Magnons, the quasiparticles of spin wave excitation, carry spin angular momentum and magnetic moments. They can exhibit topological characteristics after introducing interactions such as the Dzyaloshinskii-Moriya interaction (DMI) \cite{DZYALOSHINSKY1958241}, an antisymmetric exchange that originates from spin-orbit coupling and arises when inversion symmetry of the system is broken. In 2011, important studies predicted that a non-zero Berry curvature in magnon Chern insulators could endow magnons with orbital angular momentum, which arises from the rotational motions of the magnon wave packet in two distinct ways (see Fig. \ref{two rotations}): self-rotation and motion along the boundary of the sample (edge current) \cite{PhysRevLett.106.197202}. Recently, it has been proposed that this orbital angular momentum, induced by the Berry curvature from the DMI, can contribute to the EdH effect. This concept is referred to as the EdH effect of topological magnons, where the gyromagnetic ratio is defined as $\gamma_{m}=\frac{L}{\Delta M}$, with $L$ being the topological angular momentum and ${\Delta M}$ the magnon moment \cite{PhysRevResearch.3.023248}.

The EdH effect of topological magnons has been investigated in various lattice structures, including the square-octagon, 
honeycomb, square-hexagon-octagon, and kagome lattices, using the topological magnon formalism \cite{PhysRevResearch.3.023248, PhysRevB.108.144407}. At suitable temperatures, the square-octagon and the square-hexagon-octagon lattices display notable EdH effects. Furthermore, researchers also computationally predicts real materials where the EdH effect of topological magnons may be observed, such as $\rm{CrI_{3}}$ and Cu(1,3-bdc). The EdH effect of topological magnons signifies a direct combination of magnetism and topology. When considering magneto-mechanical coupling on this basis, it becomes apparent that the traditional EdH effect can also arise. To probe this phenomenon, the authors propose an experimental design (see Fig. \ref{exp}) \cite{PhysRevResearch.3.023248}: a disk-shaped sample is suspended via equipment, exposed to an external heat bath with a temperature gradient, and subjected to an external magnetic field $h$. By varying the temperature or the magnetic field, the disk will rotate, and its rotation angle can be measured by a mirror mounted on its axle, which deflects the incident light. During this process, both spin angular momentum and orbital angular momentum of magnons are changed; however, current methods are unable to distinguish between these two contributions. In addition to magnons, other types of magnetic excitations, such as triplons, may also influence angular momentum transfer and exhibit an analogous EdH effect.

\textit{EdH effect in Magnetic skyrmions.} Magnetic skyrmions are topologically protected domains with vortex magnetic structure, characterized by the topological number $N=\frac{1}{4\pi}\int \bm{m}\cdot(\frac{\partial \bm{m}}{\partial x}\times\frac{\partial \bm{m}}{\partial y}) dxdy $, where $\bm{m}=\bm{m(\bm{r})}$ represents the normalized magnetization at position $\bm{r}$. With advantages such as easy manipulation, small size, and high driving speed, magnetic skyrmions are considered ideal information carriers in future storage and logic computing devices. However, achieving precise manipulation of skyrmions remains a key challenge for practical implementation. We find that the magnetic angular momentum of skyrmions, $\bm{M}=\int \bm{m(\bm{r})} dr^3$, is altered when a skyrmion is created, annihilated, or undergoes a change in size. If this angular momentum is transferred to the lattice, it can lead to the emergence of the EdH effect. Notably, owing to symmetry, the magnetic angular momentum of skyrmions may remain unchanged when their structures are modified. This indicates skyrmions in two-dimensional materials are suitable candidates for implementing the EdH effect. Currently, the regulation methods for skyrmions primarily rely on external factors, such as magnetic field, electric field, and temperature variations. If one uses the EdH effect to regulate skyrmions through internal angular momentum exchange, it would undoubtedly provide a broader prospect for designing skyrmion-based devices.

To summarize, the EdH effect in topological systems connects the fields of topology, magnetism, and mechanics, offering a new approach to probing topology through mechanical responses beyond existing thermal and spin-based approaches. This represents an exciting new research frontier in magnonics.

\textit{Chirality-selective magnon-phonon coupling.} Phonons typically demonstrate linear polarization with zero angular momentum. However, in systems that break time-reversal or space-inversion symmetry, phonons can acquire non-zero angular momentum. A typical example is the nonmagnetic hexagonal lattice, where phonons present circular polarization and carry angular momentum at high-symmetry points. Furthermore, these polarized phonons possess well-defined chirality, an asymmetric property described as left- or right-handed behavior.

Phonons with non-zero angular momentum can couple with magnons via exchanging angular momentum. This coupling essentially originates from the spin-orbit interactions and manifests at a macroscopic level as a phenomenon where magnetic anisotropy is affected by atomic distances and lattice deformations \cite{doi:10.7566/JPSJ.93.034708}. The magnetic anisotropy can be the DMI, anisotropic exchange interactions, or single-ion anisotropy. It has been proposed theoretically that such magnon-phonon couplings can induce chiral phonons and tunable topological properties in both ferromagnets \cite{PhysRevLett.133.246604} and antiferromagnets \cite{PhysRevB.105.L100402}. Strong magnon-phonon coupling has been found in various materials, leading to intriguing properties. For example, in the multiferroic compound $\rm{Fe_{2}Mo_{3}O_{8}}$, a new quasiparticle called the topological magnon polaron has been detected \cite{Bao2023},  emerging from the hybridization of magnons and phonons. In $\rm{FePSe_{3}}$, which has a zigzag-type antiferromagnetic ground state, the magnon-phonon coupling exhibits chiral selectivity, occurring exclusively when the angular momenta of magnons and phonons are aligned \cite{Cui2023}. Moreover, apart from chirality, Rayleigh-type surface acoustic wave (SAW) also exhibits non-reciprocity during transmission. When coupled to the magnetization, SAWs propagating in opposite directions yield contrasting effects on magnetization precession -- 
 suppressing it in one direction
 while enhancing it in the other \cite{doi:10.1126/sciadv.abc5648}. In turn, the magnetization dynamics preferentially pumps the SAWs to the left or right, depending on the polarization of the precession direction \cite{ PhysRevLett.125.077203, PhysRevB.102.134417, 10.1063/5.0235303}.

In addition to angular momentum, phonons are suggested to possess magnetic moments \cite{Cheng2020} on scale of the Bohr magnetic moment $\mu_{B}$, which exceeds the magnetization produced by time-varying electric polarization \cite{PhysRevMaterials.3.064405, PhysRevMaterials.1.014401}, $\bm{M}\propto \bm{P}\times{\partial}_{t}\bm{P}$. The latter is caused by ionic motions and is on the order of the nuclear magneton, approximately $10^{-4}\mu_{\rm{B}}$ to $10^{-3}\mu_{\rm{B}}$. To date, phonon magnetic moments have been identified in paramagnetic \cite{doi:10.1126/science.adi9601}, ferromagnetic \cite{Wu2023}, and non-magnetic materials \cite{Basini2024}, with various explanations proposed for this phenomenon. In paramagnetic or magnetically ordered systems, it has been suggested that phonons can acquire magnetic moments through coupling to magnons; in contrast, in non-magnetic systems, this effect is primarily attributed to electronic excitations or electron-phonon coupling. More recently, a microscopic model based on orbit-lattice coupling has been developed to explain phonon magnetic moments in paramagnetic and magnetic materials \cite{PhysRevB.110.094401}. In this model, a degenerate chiral phonon mode couples to a degenerate orbital transition, forming two hybridized branches: one with a major phonon contribution and another with a major orbital contribution. Consequently, the phonon mode acquires a part of the $g$ factor from the orbital transition, which is several orders of magnitude larger than its own \cite{PhysRevB.110.094401}. This result implies the possibility of a unified theory regarding the origin of phonon magnetic moments.

The mechanism underlying the phonon thermal Hall effect has been inconclusive since its discovery in 2005, as phonons are electrically neutral particles that must be modulated by magnetic fields with the help of interaction with electrons or spins. The proposed mechanisms can be categorized into three groups: Raman-type spin-phonon interaction \cite{PhysRevLett.96.155901}, Berry curvature \cite{PhysRevLett.105.225901}, and phonon scattering \cite{PhysRevLett.113.265901}. However, these mechanisms do not account for all observed phonon Hall effects, particularly in what are considered ``trivial'' systems, such as nonmagnetic, paramagnetic, and symmetric materials \cite{jin2024discoveryuniversalphononthermal}. We infer that for these systems there is a complex interplay between phonons and electrons, in which the charge fluctuations are nonnegligible. Notably, the similarities in materials and mechanisms involved in the phonon magnetic moment and the phonon thermal Hall effect suggest a potential connection between these phenomena. Could the understanding of phonon magnetic moments provide a new perspective for explaining the phonon thermal Hall effect?

A large phonon magnetic moment greatly challenges the conventional view that phonons cannot directly respond to an electromagnetic field, positioning phonons as promising candidates for device design akin to electrons and magnons. This shift in understanding allows researchers to explore phonon degrees of freedom from a fresh perspective. Moreover, if the phonon magnetic moment is clearly understood, then the ``magic'' phonon which has both angular momentum and magnetic moment, can itself give rise to the EdH effect.

\section*{ACKNOWLEDGEMENT}
This work was supported by the National Key Research and 
Development Program of China (2022YFA1402802), the National 
Natural Sciences Foundation of China (92165204 and 12494591), 
Leading Talent Program of Guangdong Special Projects 
(201626003), Guangdong Provincial Key Laboratory of Magnetoelectric Physics and Devices (2022B1212010008), Research Center 
for Magnetoelectric Physics of Guangdong Province 
(2024B0303390001), and Guangdong Provincial Quantum Science 
Strategic Initiative (GDZX2401010).

\end{document}